\documentclass[12pt,preprint2]{aastex}
\usepackage{natbib}
\usepackage{amsmath}
\usepackage{epsfig}
\usepackage{graphicx}
\usepackage{amssymb}
\usepackage{color}
\usepackage{epstopdf}
\usepackage{amsmath}
\usepackage{amsfonts}
\usepackage{amssymb}
\newcommand{\eps}{e$^{-}/$s}
\newcommand{\nw}{nW m$^{-2}$ sr$^{-1}$}

\definecolor{orange}{rgb}{1,0.5,0}
\definecolor{purple}{rgb}{0.7,0,0.7}

\shorttitle{CIBER: the Narrow Band Spectrometer}
\shortauthors{Korngut et al. (The CIBER Collaboration)}

\begin{document}
\slugcomment{Draft version \today; }

\title{The Cosmic Infrared Background Experiment (CIBER): the Narrow Band Spectrometer}

\author{P.~M.~Korngut\altaffilmark{1,2,13}, 
  T.~Renbarger\altaffilmark{3}, T.~Arai\altaffilmark{4,5},
  J.~Battle\altaffilmark{2}, J.~Bock\altaffilmark{2,1},
  S.~W.~Brown\altaffilmark{6}, A.~Cooray\altaffilmark{7},
  V.~Hristov\altaffilmark{2}, B.~Keating\altaffilmark{3},
  M.~G.~Kim\altaffilmark{8}, 
  A.~Lanz\altaffilmark{2}, D.~H.~Lee\altaffilmark{9},
  L.~R.~Levenson\altaffilmark{2}, K.~R.~Lykke\altaffilmark{6},
  P.~Mason\altaffilmark{2}, T.~Matsumoto\altaffilmark{4,8,12},
  S.~Matsuura\altaffilmark{4}, U.~W.~Nam\altaffilmark{9},
  B.~Shultz\altaffilmark{10}, A.~W.~Smith\altaffilmark{6},
  I.~Sullivan\altaffilmark{11}, K.~Tsumura\altaffilmark{4},
  T.~Wada\altaffilmark{4}, and M.~Zemcov\altaffilmark{2,1}}

\altaffiltext{1}{Jet Propulsion Laboratory (JPL), National Aeronautics
  and Space Administration (NASA), Pasadena, CA 91109, USA}
\altaffiltext{2}{Department of Physics, California Institute of
  Technology, Pasadena, CA 91125, USA}
\altaffiltext{3}{Department of Physics, University of California, San
  Diego, San Diego, CA 92093, USA}
\altaffiltext{4}{Department of Infrared Astrophysics, Institute of
  Space and Astronautical Science (ISAS), Japan Aerospace Exploration
  Agency (JAXA), Sagamihara, Kanagawa 252-5210, Japan}
\altaffiltext{5}{Department of Physics, Graduate School of Science,
  The University of Tokyo, Tokyo 113-0033, Japan}
\altaffiltext{6}{Sensor Science Division, National Institute of
  Standards and Technology (NIST), Gaithersburg, MD 20899, USA}
\altaffiltext{7}{Center for Cosmology, University of California,
  Irvine, Irvine, CA 92697, USA}
\altaffiltext{8}{Department of Physics and Astronomy, Seoul National
  University, Seoul 151-742, Korea}
\altaffiltext{9}{Korea Astronomy and Space Science Institute (KASI),
  Daejeon 305-348, Korea}
\altaffiltext{10}{Materion Barr Precision Optics \& Thin Film
  Coatings, Westford, MA, 01886, USA}
\altaffiltext{11}{Department of Physics, The University of Washington,
  Seattle, WA 98195, USA}
\altaffiltext{12}{Institute of Astronomy and Astrophysics, Academia Sinica, Room 1314, Astronomy-Mathematics Building, National Taiwan University　No.1, Roosevelt Rd, Sec. 4 Taipei 10617, Taiwan, R.O.C.}
\altaffiltext{13}{Contact author, pkorngut@caltech.edu}

\begin{abstract}
We have developed a near-infrared spectrometer designed to measure the
absolute intensity of the Solar 854.2 nm \ion{Ca}{2} Fraunhofer line,
scattered by interplanetary dust, in the Zodiacal light spectrum.  Based on
the known equivalent line width in the Solar spectrum, this
measurement can derive the Zodiacal brightness,
testing models of the Zodiacal light based on morphology
that are used to determine the extragalactic background light in
absolute photometry measurements.  The spectrometer is based on a
simple high-resolution tipped filter placed in front of a compact
camera with wide-field refractive optics to provide the large optical
throughput and high sensitivity required for rocket-borne observations.  We
discuss the instrument requirements for an accurate measurement of
the absolute Zodiacal light brightness, the measured laboratory characterization,
and the instrument performance in flight.
\end{abstract}

\keywords{infrared: diffuse background --- instrumentation:
  spectrograph --- methods: laboratory --- space vehicles: instruments
  --- techniques: spectroscopic --- zodiacal dust}

\section{Introduction}
\label{S:intro}

\setcounter{footnote}{0}

The intensity of the infrared Extragalactic Background Light (EBL) is a cornerstone of
cosmological observations, a measure of the total radiation produced
by stellar nucleosynthesis and gravitational accretion over cosmic
history.  Accurate measurements can determine if the EBL
is consistent with the calculated surface brightness from galaxy
counts, or if an additional background component is present, e.g.,
unaccounted flux from known galaxies or missing galaxy populations.
Furthermore, light from the epoch of reionization must contribute to
the total EBL, with Lyman cutoff and Lyman-$\alpha$ features originating at $z >6$
redshifted into the near-infrared (NIR) wavelength band.  

Current direct measurements of the NIR EBL intensity are mutually inconsistent.  The dominant factor causing this discrepancy arises not from instrumental effects, but from different approaches to astrophysical foreground removal.
 In the NIR, all astronomical observations taken 1~AU from the Sun are completely dominated by Zodiacal Light (ZL).  ZL originates from
interplanetary dust scattering sunlight at optical and near-infrared
wavelengths, and emitting thermal radiation at mid-infrared to
far-infrared wavelengths.  In the early 1990's, the Diffuse Infrared Background Experiment (DIRBE) produced absolute
measurements of the astrophysical sky brightness in multiple bands
over a wide range of Solar elongation angles \citep{Hauser1998}.  A
model of ZL based on morphology derived from annual modulation of the signal measured by DIRBE was developed for
subtracting the ZL foreground in order to assess the EBL \citep{Kelsall98}.  Applying this model has led to high EBL estimates \citep{Cambresy2001,Dwek1998,Matsumoto05}. 
Zodiacal dust models based on morphology alone are not unique, and a
separate ZL model was subsequently developed based on the assumption that
the mid-infrared sky brightness observed by DIRBE is entirely due to ZL
\citep{Wright2001}.  Implementing this foreground treatment led to a significant decrease in estimated EBL levels \citep{Levenson2007}. \citet{Dwek05} make the important observation that a 23\% underestimate in ZL intensity in the \citet{Kelsall98} model would account for the entire observed NIR EBL excess, as the residual EBL spectrum displays the same color as ZL within the statistical uncertainties. An independent test of the ZL contribution to total sky brightness is needed.

Solar Fraunhofer lines scattered by interplanetary dust provide an alternative method of estimating ZL brightness.
As the equivalent widths of Fraunhofer lines are accurately known in the
Solar spectrum, a measurement of the line in scattered light is
directly related to the ZL continuum brightness.  Empirically, the
measured equivalent width of Fraunhofer lines in ZL matches the Solar
spectrum to $< 2\%$ \citep{Beggs1964}.  Early measurements by \citet{Dube77} were the first to estimate ZL from its Fraunhofer signature. \citet{BernsteinZL}
attempted to measure the optical EBL using a combination of absolute
sky brightness measurements with the Hubble Space Telescope and
Fraunhofer line measurements with a ground-based telescope.  This pioneering measurement was complicated by
atmospheric airglow emission, but also by systematic uncertainties
from atmospheric scattering, ground scattering, and stray light
\citep{Mattila2003}.  
\citet{Mattila2011} have developed a technique to
remove ZL by measuring the difference in brightness between a
high-latitude dark nebula and its surrounding area, using Fraunhofer
line measurements to remove diffuse Galactic light from starlight
scattered by interstellar dust. They have applied this technique to estimate the EBL at 400~nm and provide an upper limit at 520~nm.    
A recent re-analysis of Pioneer 10
and 11 optical data at 440~nm and 640~nm from outside the Zodiacal cloud
\citep{Matsuoka2011} also derive an EBL that is consistent with
\citet{Mattila2011}.  These measurements, done with a model-independent ZL foreground estimation, suggest a total EBL consistent with the integrated light from galaxies in the optical.

In this paper, we describe the design and implementation of a Narrow Band Spectrometer (NBS) aboard the Cosmic Infrared Background Experiment (CIBER) sounding rocket payload \citep[see][]{Zemcov11}.  The instrument is optimized to measure ZL intensity by observing a single Fraunhofer line at 854.2 nm caused by Solar \ion{Ca}{2} absorption, and work in synergy with a low-resolution spectrometer (LRS) \citep[see][]{Tsumura11} to estimate the mean intensity of the EBL in the NIR.  The LRS will measure the spectrum of the
ZL (e.g. \citealt{Tsumura10}) from 0.75 to $2.1 \, \mu$m
so that a determination of the ZL brightness at $854.2 \,$nm can be
directly compared with predictions from ZL models based on the DIRBE
1.25 and $2.2 \, \mu$m bands, where the EBL discrepancies are the largest. The CIBER payload also contains two wide field imaging camera \citep{Bock2012}, designed to probe the EBL through its spatial fluctuation characteristics.

\section{Instrument Description}
\label{S:instrument}

We have chosen to measure the 854.2 nm line,
the strongest of the \ion{Ca}{2} triplet lines at 849.8, 854.2 and $866.2 \,$nm.
We selected a long-wavelength Fraunhofer line for comparison with the
DIRBE/Kelsall model at 1.25 and $2.2 \, \mu$m, and the $854.2 \,$nm
line is the brightest non-hydrogenic Fraunhofer line, with an equivalent width
(EW) of $ 0.37 \,$nm \citep{Allen1976}, at wavelengths longer than $400
\,$nm.  We avoided hydrogen lines, which have a
contribution from recombination line emission from the interstellar warm
ionized medium \citep{Martin1988}.  The narrow-band spectrometer
(NBS) uses a tipped filter spectrometer with high-throughput
refractive optics \citep{Eather1969}. 
Our goal is to make a percent-level determination of the ZL intensity,
in comparison to the current 23~\% error needed to explain the discrepancy
between the excess EBL and galaxy counts \citep{Dwek05}.

\subsection{ZL Detection}
\label{sS:requirements}

The NBS measures the absolute sky brightness on and off a known Fraunhofer
line, producing a photocurrent
\begin{equation}
\label{eq:Ilambda}
i(\lambda) = \frac{\eta A \Omega}{h c} \int \lambda^{\prime} I_{\lambda^{\prime}}
T(\lambda,\lambda^{\prime}) d\lambda^{\prime} \mathrm{\hspace{0.5cm}
  [e} ^{-} \mathrm{/s]},
\end{equation}
where $h$ and $c$ are Planck's constant and the speed of light,
respectively, $\eta$ is the peak efficiency, $A \Omega$ is the
detector throughput, $I_{\lambda^{\prime}}$ is the sky surface
brightness, and $T(\lambda,\lambda^{\prime})$ is the normalized
instrument response function centered at wavelength $\lambda$.  High
instrument sensitivity is needed in order to make precise measurements
during the short observation time available in a sounding rocket
flight.  In the simplified case where the sky consists of two
brightness components, the ZL and the EBL, and the Fraunhofer line is unresolved, the photo currents
measured on and off the line are
\begin{equation}
\label{eq:Ion}
i_{\mathrm{on}} \simeq \frac{\eta A \Omega}{h c} (\lambda
I_{\lambda,\mathrm{ZL}} \Delta \lambda - \lambda
I_{\lambda,\mathrm{ZL}} W + \lambda I_{\lambda,\mathrm{EBL}}
\Delta \lambda),
\end{equation}
\begin{equation}
\label{eq:Ioff}
i_{\mathrm{off}} \simeq \frac{\eta A \Omega}{h c} (\lambda
I_{\lambda,\mathrm{ZL}} \Delta \lambda + \lambda I_{\lambda,\mathrm{EBL}}
\Delta \lambda), \mathrm{ and}
\end{equation}
\begin{equation}
\label{eq:Idiff}
i_{\mathrm{off}} - i_{\mathrm{on}} = \frac{\eta A \Omega}{h c} \lambda
I_{\lambda,\mathrm{ZL}} W,
\end{equation}
where $\lambda I_{\lambda,\mathrm{ZL}}$ and $\lambda
I_{\lambda,\mathrm{EBL}}$ are the ZL and EBL continuum brightness off
the line, $W$ is the equivalent width of the Fraunhofer line, and the
spectral resolution is
\begin{equation}
\label{eq:specres}
\Delta \lambda = \int T(\lambda,\lambda^{\prime}) d \lambda^{\prime}.
\end{equation}
Thus the estimated ZL local continuum brightness is simply
\begin{equation}
\label{eq:contbrightness}
\lambda I_{\lambda,\mathrm{ZL}} = \frac{hc}{W \eta A \Omega}
(i_{\mathrm{off}} - i_{\mathrm{on}}),
\end{equation}
determined with an uncertainty given by
\begin{equation}
\label{eq:contacc}
\sigma _{\lambda I_{\lambda,\mathrm{ZL}}} = \frac{hc}{W \eta A \Omega}
(\sigma i_{\mathrm{off}}^{2} + \sigma i_{\mathrm{on}}^{2})^{1/2},
\end{equation}
where $\sigma i_{\mathrm{off}}$ and $\sigma i_{\mathrm{on}}$ are the
instrumental uncertainties in the on and off line detector currents.
If the noise on detector current scales as the number of pixels
$N$ on and off the line, the ZL sensitivity is given by
\begin{equation}
\label{eq:ZLsens}
\sigma _{\lambda I_{\lambda,\mathrm{ZL}}} = \frac{hc}{W \eta A \Omega}
\sigma i (2/N)^{1/2},
\end{equation}
where $\sigma_{i}$ is the per pixel current noise.  Thus to optimize
surface brightness sensitivity, we need to maximize the pixel
throughput $A \Omega$, and incorporate as many pixels on and off the
line as possible. 

In addition to the measurement errors, $W$ technically contains  uncertainty itself, dominated by the stability of the Fraunhofer line profile over time.  To quantify this contribution, we obtained publicly available high resolution spectra from the SOLIS\footnote{http://solis.nso.edu/iss/} instrument which sampled the \ion{Ca}{2} line daily over the course of several years (2008 to 2012).  The equivalent width of the line measured between 853.55~nm and 854.80~nm was observed to vary at the 0.5\% level over this time period, and is therefore taken to be a highly sub-dominant source of error in ZL measurement using this technique. 

In addition to raw sensitivity, we require sufficient spatial
resolution to remove the foreground contribution from stars.  As the
\ion{Ca}{2} Fraunhofer line is expected to have a comparable depth in the
interstellar radiation field as in the Solar spectrum \citep{Mattila1980},
residual starlight will add to the inferred ZL intensity in Eq.~5.  We
have chosen a pixel resolution of $2 \times 2$ arcmin in order to
mask stars down to $S < 14$ Vega magnitude using positions and fluxes from an external catalog.

\subsection{Optical Design}
\label{sS:opticaldesign}

\begin{figure*}[hbtp]
\begin{center}
\includegraphics[width=150mm]{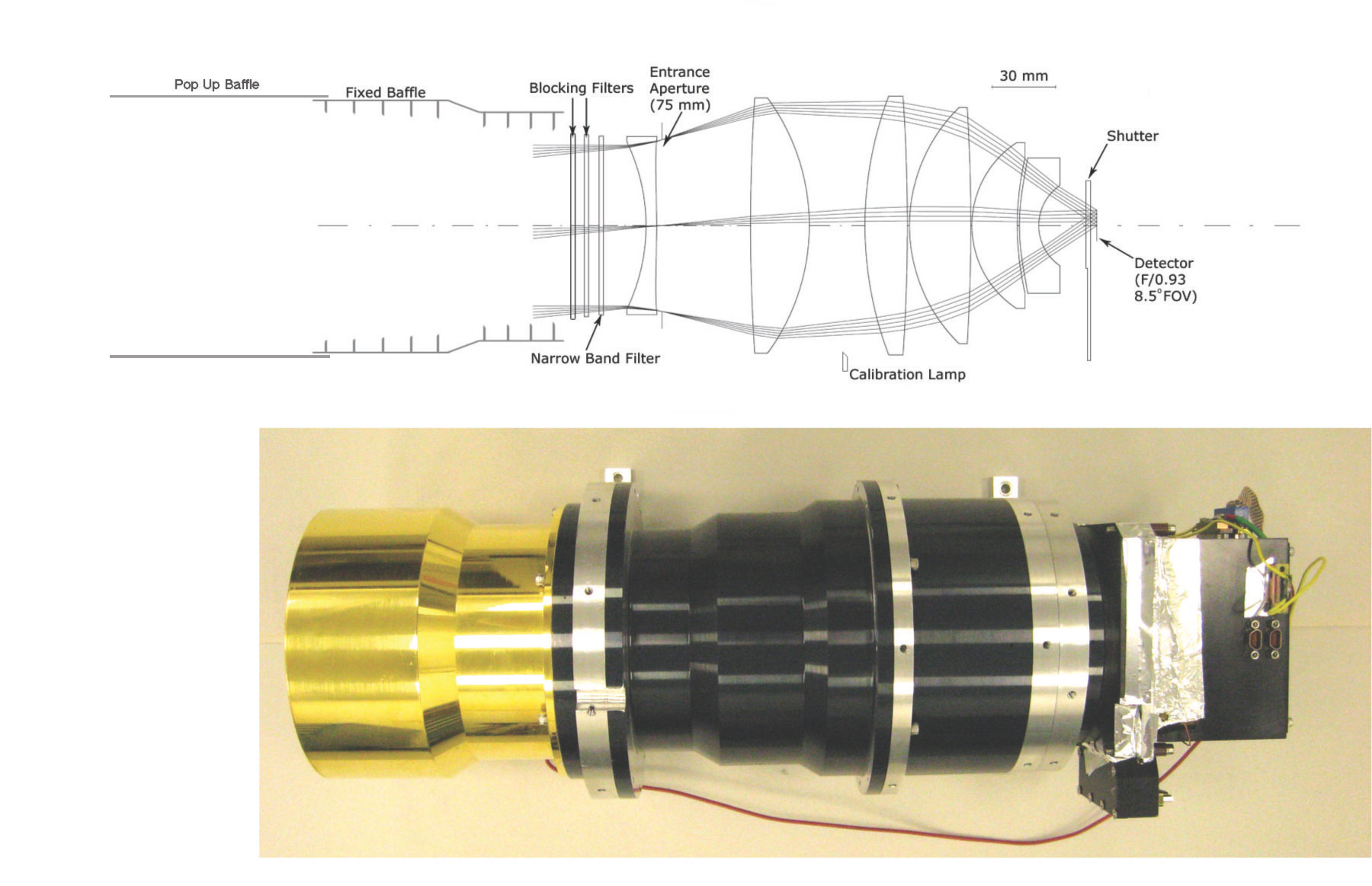}
\caption{Ray trace and photo of the NBS optics.  The spectrometer uses
  a fixed interference filter combined with 2 blocking filters placed
  at the entrance aperture of a wide-field camera (see Table
  \ref{tab:optics}\ for specifications).  The entire optical assembly
  and focal plane assembly is cooled to $77 \,$K in operation.  The
  ray trace shows an extended ``pop up" baffle tube used after the first flight
  that is not presented in the photograph.}
\end{center}
\label{fig:nbs_raytrace}
\end{figure*}

The NBS instrument uses a narrow-band interference filter placed at
the aperture of an imaging camera with high-throughput refractive
optics.  This \textquoteleft tilted-filter photometer\textquoteright \space provides
an intrinsic sensitivity advantage by having higher throughput
than spectrometers using a prism or grating (\citealt{Jacquinot1954},
\citealt{Eather1969}).  The design of the narrow-band spectrometer
optics is shown in Figure \ref{fig:nbs_raytrace}, and detailed
specifications are summarized in Table \ref{tab:optics}.

The narrow-band interference filter was fabricated by Materion Barr
 Precision Optics \& Thin Film Coatings\footnote{{\url http://materion.com/}} to customized specifications from the science team. 
The filter is tipped by $2^{\circ}$ from normal in order to
optimize sensitivity, based on the number of detectors on and off the
line.  The narrow-band filter is combined with two filters that
provide out-of-band spectral blocking.

The wide-field camera is a refractive design with six
lenses developed and manufactured by
 the Genesia Corporation, Japan\footnote{{\url http://www.genesia.co.jp/}}
based on the cryogenic refraction measurements discussed in
\citet{Yamamuro2006}.  
The lenses are manufactured from fused
silica and are spherical except for the back surface of the entrance
lens and the front surface of the last lens, which are
$6^{\mathrm{th}}$ order aspheres.  The lenses are anti-reflection
coated for $< 0.2 \%$ reflectance per surface at $854.2 \,$nm.

A $256 \times 256$ PICNIC HgCdTe detector array
manufactured by Teledyne Scientific \& Imaging\footnote{{\url http://www.teledyne-si.com/}} is placed at
the $f/0.92$ focus of the camera, yielding a $2^{\prime} \times 2^{\prime}$
pixel scale and an $8.5^{\circ} \times 8.5^{\circ}$ field of view
on the sky.  A mechanical shutter and calibration lamp are
also installed in the NBS instrument (see \citet{Zemcov11} for
detailed specifications of these subsystems).  The cryogenic shutter is
placed between the optics and the detector array in order to monitor
array dark current.  
A calibration lamp assembly feeds light from a light emitting diode
into the optical chain with an optical fiber between the third and
fourth lenses.  The lamp is used as a transfer standard to track any
changes in response in flight compared with the laboratory
calibration.

\begin{table*}
\centering
\caption{NBS Component Parameters.}
\begin{tabular}{lc}
\hline
Optics & wide field refractive camera with 6 lenses \\

Filters & 2 blocking filters and 1 tipped narrow-band filter. \\

Detector & $256 \times 256$ MBE HgCdTe PICNIC \\

Aperture & $75 \,$mm \\

$f/\#$ & 0.92 \\

Pixel size & $2 \times 2\,$arcmin \\

Field of view & $8.5^{\circ} \times 8.5^{\circ}$ \\

Filter central wavelength at 295 K & $855.57 \,$nm \\

Filter central wavelength at 77 K & $854.55 \,$nm \\

Free spectral range at 77 K & $852.0 - 854.5 \,$nm \\

Filter resolution $\lambda / \Delta \lambda^{\mathrm{a}}$ & $1120$ \\

Filter tip angle & $2.0^{\circ}$ \\

Lens efficiency & $0.98$ \\

ED 541$^{\mathrm{b}}$ narrow-band filter efficiency & $0.84$ \\

ED 536 blocking filter efficiency & $0.66$ \\

ED 686 blocking filter efficiency & $0.83$ \\

Detector QE$^{\mathrm{c}}$ & $0.83$ \\

Total QE & $0.37$ \\

Correlated double sample noise & $28 \,$e$^{-}$ \\

Calculated off-line responsivity$^{\mathrm{d}}$ & $529~$
 nW m$^{-2}$ sr$^{-1}$ /~ e$^{-}$ s$^{-1}$  \\

\hline
\multicolumn{2}{l}{$^{\mathrm{a}}$ $\Delta \lambda$ is the integral
  width $\Delta \lambda = \int T(\lambda^{\prime}) d \lambda^{\prime}$.} \\
\multicolumn{2}{l}{$^{\mathrm{b}}$ ED NNN is a manufacturer's part number.} \\
\multicolumn{2}{l}{$^{\mathrm{c}}$ Measured at $2.2 \, \mu$m.} \\
\multicolumn{2}{l}{$^{\mathrm{d}}$ Responsivity given by $\eta \Delta
  \lambda A \Omega / h c$ in Eq.~\ref{eq:Ioff}.} \\

\end{tabular}
\label{tab:optics}
\end{table*}

\subsection{Filter Design}
\label{sS:filterdesign}

\begin{figure}[hbtp]
\begin{center}
\includegraphics[width=65mm]{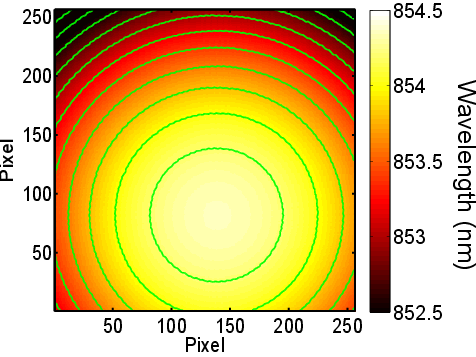}
\end{center}
\caption{Instrument wavelength map calculated from Equation
  \ref{eq:NBSlaw} using the parameters listed in Table
  \ref{tab:optics}.  The colorscale shows the peak wavelength response
  of each pixel in the $256 \times 256$ detector array, with the
  contours at intervals of $\Delta \lambda = 0.2 \,$nm.  }
\label{fig:ideallambda}
\end{figure}

The spectrometer disperses radiation by shifting the effective filter
bandpass by the change in incident angle over the field of view (FOV)
according to
\begin{equation}
\label{eq:NBSlaw}
\lambda = \lambda_{0} \cos (\theta_{i}),
\end{equation}
where $\lambda_{0}$ is the response at normal incidence and the
internal angle $\theta_{i}$ is related to the incident external angle
$\theta_{e}$ by Snell's law $\sin (\theta_{i}) = \sin (\theta_{e}) /
n$.  Figure \ref{fig:ideallambda} shows the calculated peak wavelength
response of pixels across the array using the parameters listed in
Table \ref{tab:optics}, and exhibits the characteristic shift given
by Equation \ref{eq:NBSlaw}.  The $8.5^{\circ} \times 8.5^{\circ}$
field of view combined with the $2^{\circ}$ filter tip modulates the
wavelength response from $\lambda - \lambda_{0} = 0$ at normal filter
incidence to $\lambda - \lambda_{0} = 0.003 \lambda_{0}$ at the
extreme corners.  The divergence-limited resolution
(e.g.~\citealt{Bock1994}) varies from $\lambda/\Delta \lambda = 3.6
\times 10^{7}$ at the array center to $2.8 \times 10^{4}$, at the minimum
wavelength at the array corner.  The spectral range of $\lambda/\Delta \lambda =
1/0.003 = 330$ is sufficient to sample the ZL spectrum both on and
off the \ion{Ca}{2} $854.2 \,$nm Fraunhofer line, which itself has a width of
$\lambda / \Delta \lambda_{\mathrm{FWHM}} \approx 4000$.

The instrument is insensitive to the choice of spectral resolution to
first order (see Equation \ref{eq:contacc}).  However, the
resolution does vary the number of pixels on and off the line.  Higher
resolution makes $N_{\mathrm{on}} < N_{\mathrm{off}}$, slowly reducing
sensitivity but also providing for a larger spectral contrast to ZL
according to
\begin{equation}
\label{eq:contrast}
\frac{i_{\mathrm{off}} {-} i_{\mathrm{on}}}{i_{\mathrm{off}}} =
  \frac{W}{\Delta \lambda}.
\end{equation}

\begin{figure}[hbtp]
\begin{center}
\includegraphics[width=65mm]{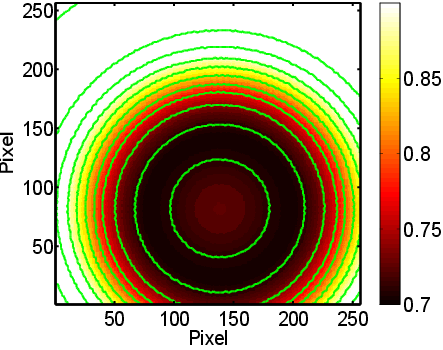}
\end{center}
\caption{The convolved Solar spectrum with the calculated
  spectral response over the field of view of the NBS using the
  instrument parameters given in Table \ref{tab:optics}.  We plot the
  ZL intensity normalized to unity at the ZL continuum level.  The
  contours show annuli of constant brightness and are spaced at
  intervals of $0.03$.  The drop in signal around the maximum
  wavelength region is due to the $854.2 \,$nm \ion{Ca}{2} line.
  Annular averages centered on the filter response pattern give a
  known profile from the Solar spectrum that is directly proportional
  to the ZL intensity.}
\label{fig:solarimage}
\end{figure}

\begin{figure}[htb]
\begin{center}
\includegraphics[width=65mm]{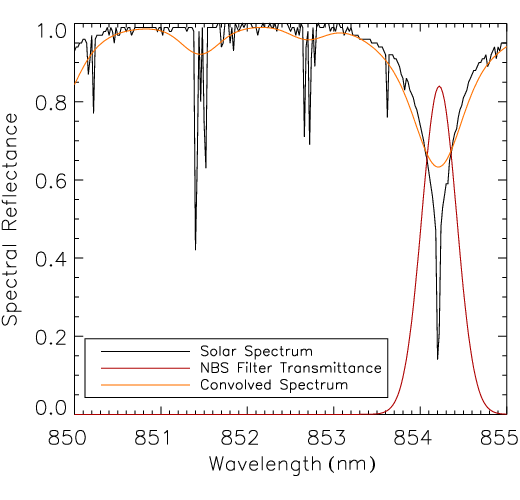}
\end{center}
\caption{Comparison of the Solar spectrum \citep{Debouille1990} near
  the \ion{Ca}{2} $854.22 \,$nm line, the spectral response of the NBS
  filter, and the convolution of the filter with the Solar spectrum.
  The NBS measures the convolved spectrum from $852.0$ to $854.5\,$nm.
  The NBS filter has been normalized to the peak transmittance listed
  in Table \ref{tab:optics}. } 
\label{fig:solartheory}
\end{figure}

The actual response to ZL over the FOV, given the finite resolution
and the finite width of the Fraunhofer line, is the
convolution of the instrument response with the Solar spectrum (see
Equation \ref{eq:Ilambda}).  The spectral contrast on and off the
\ion{Ca}{2} line is $26 \,$\%, giving the characteristic pattern shown
in the noiseless instrument simulation in Figure \ref{fig:solarimage}.
Using a simulator we investigated a range of resolution and
filter tips, balancing sensitivity and contrast, and determined
that $\lambda / \Delta \lambda \approx 1000$ was a good choice.  Since
the wavelength response of the pixels varies with radius from the
maximum wavelength, data from the NBS measures the convolved image
shown in Figure \ref{fig:solarimage}, from which the absolute ZL
brightness can be determined.  The spectral response convolved with
the Solar spectrum is shown in Figure \ref{fig:solartheory}.

\subsection{Detectors and Electronics}
\label{sS:Electronics}

The NBS uses a $256 \times 256$ HgCdTe detector array
employing a PICNIC read out integrated circuit (ROIC).  A molecular beam epitaxy (MBE) process is used to provide short wavelength response compared with HgCdTe arrays fabricated on sapphire substrates (M.~Farris, private communication).  The detectors have a $40 \, \mu$m
pixel pitch, giving a large area optimal surface brightness sensitivity.  A
custom electronics chain is used to address pixels on the multiplexer
and read out and digitize the detector signals, described in detail in
\citet{Zemcov11}.  The detector bias and reset voltages are trimmed as
in \citet{Lee2010}.  We do not use the built-in buffer amplifier
circuitry on the ROIC, which emits photons and increases the detector dark
current, in favor of external junction field-effect transistor (JFET) source
followers \citep{Hodapp1996}.  Since the same detectors are used in the both the NBS and LRS, we refer the reader to \citet{Tsumura11} for a detailed discussion of the
electrical and responsivity properties of the CIBER PICNIC arrays.

\section{Instrument Assembly and Laboratory Testing}
\label{S:testing}

We tested the instrument performance in the laboratory prior to
flight.  The payload section is designed to accommodate several test
configurations \citep{Zemcov11}: 
\begin{enumerate}

\item Using a vacuum lid with fused quartz
windows the NBS can view external sources to measure focus, wavelength
response, and absolute calibration using the  SIRCUS (Spectral Irradiance and radiance Responsivity Calibrations using Uniform Sources) facility
\citep{Brown06}.

\item Using a vacuum box the NBS views an integrating
sphere without intervening windows to measure the flat field.

\item Using the flight shutter door equipped with a radiatively cooled inner
shield we can measure the noise properties and dark current under dark
conditions.

\end{enumerate}

\subsection{Focus testing}
\label{sS:focus}

We measure the optical focus with the CIBER cryostat outfitted with
two sets of windows: warm vacuum windows installed at the bulkhead,
and cold windows operating at $\approx 150 \,$K to intercept thermal
radiation from the warm window and bulkhead.  The NBS vacuum window is
anti-reflection (AR) coated to eliminate reflections between the
window and the NBS filters, though the cold window was not.  The
transmittance of the warm and cold windows at $850 \,$nm are $0.94$
and $0.92$, respectively.

We introduced a collimated beam using a $254 \,$mm diameter $f/3.6$
Newtonian telescope with white light shining through a $50 \, \mu$m
diameter pinhole located at the telescope focus.  The best focus of
the Newtonian telescope was first determined using an auto-collimation
technique with a flat mirror.  We then illuminated the NBS entrance
aperture with the collimated telescope beam, and the pinhole was
stepped in $2.5 \,$mm increments over a $50 \,$mm range about
the telescope focus.  The best focus of the NBS was measured and the
displacement from the array position, $\Delta x_{2}$, was calculated using
\begin{equation}
\label{eq:focus}
\frac{f_{1}^{2}}{\Delta x_{1}} = \frac{f_{2}^{2}}{\Delta x_{2}},
\end{equation}
where $f_{1}$ and $f_{2}$ are the focal lengths and $\Delta x_{1}$ and
$\Delta x_{2}$ are the displacement from best focus of the collimating
telescope and the NBS, respectively.  The range of measurement allows a
$300 \, \mu$m through focus to be probed in increments of $15 \, \mu$m
from NBS focus.  We measured focus near the center of the array, at each
of the four corners of the array, and then repeated the focus test at
the array center for redundancy.

Once a focus test is completed, the position of the focal plane was
determined by plotting the full-width half-max spot size as a function
of pinhole position.  A quadratic polynomial was fit to the spot size as a function of focus position, and the minimum of this fit was taken to be the NBS array
position relative to ideal focus.  This program allowed us to focus the NBS to
within $10{-}20 \, \mu$m, well within the $f l_{p} \approx 40 \, \mu$m depth of
focus of the NBS, where $f = 0.92$ is the f/\# of the NBS and $l_{p} = 40~\mu m $
is the pixel size.

\subsection{Wavelength Calibration}
\label{sS:wavelength}

\begin{figure}[hbtp]
\begin{center}
\includegraphics[width=65mm]{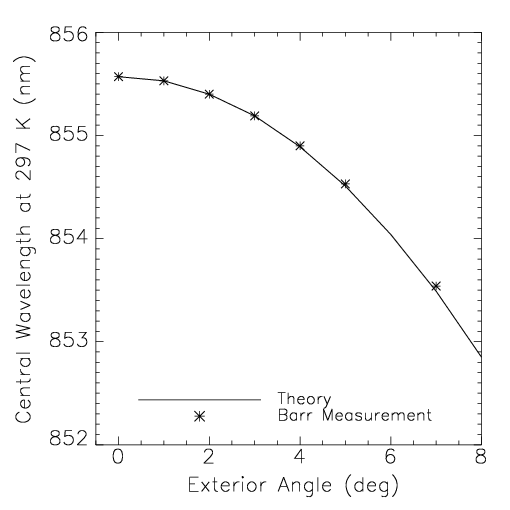}
\end{center}
\caption{Measured peak wavelength with angle of incidence at room
  temperature, fit to an index of 1.75 and $\lambda_{0} = 852.57
  \,$nm.  The theory curve follows Equation \ref{eq:NBSlaw}.  The
  wavelength response at normal incidence is a function of temperature
  so that $\lambda_{0}$ at $77 \,$K is $854.55 \,$nm.}
\label{fig:peak_wav_angle}
\end{figure}

We developed a system of filters manufactured by Barr Associates
consisting of a narrow-band filter and two blocking filters.  Due to
the precise tuning requirements, Barr measured the change in response
($\Delta \lambda = -1.0 \,$nm) on cooling to $77 \,$K on a witness
filter, and calibrated the shift in response versus incident angle at
room temperature.  The change in response is well fit to a material
index of $1.75$, as shown in Figure \ref{fig:peak_wav_angle}.  The
narrow-band filter gives a peak response of $855.57 \,$nm with an integral
width of $0.76 \,$nm at room temperature, which changes to $854.55 \,$nm when
cooled to $77 \,$K.  Barr measured the transmittance in 7 positions from
the center to the edge at room temperature, and found an average peak
transmittance of 84\% and that the peak wavelength varies by $0.34
\,$nm, maximum to minimum, at different positions across the filter.

We calibrated the NBS as a function of wavelength using the SIRCUS lasers and an absolutely calibrated reference radiometer.  
Because the laser SIRCUS Ti-Sapphire line-width is essentially zero, the narrow band pass of the NBS is fully resolved.
 Etalons placed in the laser cavity provided wavelength stability and fine tuning.  We measured wavelength response to an uncertainty of approximately 0.001~nm referenced to a calibrated wavemeter
incorporating a stabilized HeNe laser reference. 
The absolute radiance of the integrating sphere, viewed by the NBS, was inferred using a monitor detector separately calibrated to an absolutely calibrated radiometer. This two-step bootstrapping approach is necessary due to the dynamic range mismatch between the radiometer and the NBS.
In order to achieve sufficient signal to noise on the monitor detector, the monitor was located on a smaller fiber-fed sphere injecting light into the larger sphere viewed by the NBS. The relatively spectrally flat ratio of the large sphere radiance to the monitor signal was measured at a few wavelengths and interpolated to the NBS wavelengths. The sphere radiance was stabilized using the monitor signal and a liquid crystal laser intensity modulator. The monitor signal was recorded throughout the NBS calibration measurements.  

\begin{figure}[htb]
\centering
\includegraphics[width=65mm]{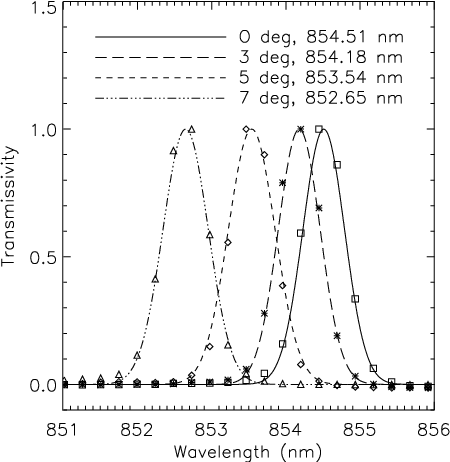}
\caption{The wavelength response for individual pixels sampled at four
  discrete angles from the maximum wavelength response of the NBS.
  The points show the measurements for four pixels,
  starting at the peak wavelength of $854.51 \,$nm and moving
  shortward.  The measurements use the SIRCUS facility, taken in wavelength
  steps of $\Delta \lambda = 0.1 \,$nm.  The curves show the fit of a Gaussian, yielding a central wavelength for each pixel.}
\label{fig:laserfilters}
\end{figure}

\begin{figure}[htb]
\centering
\includegraphics[width=65mm]{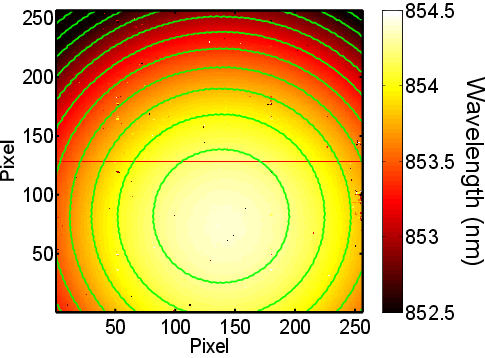}
\caption{Measured spectral response over the instrument field of view.
  The spectral response was measured using a high-resolution tunable
  laser injected into an integrating sphere with an exit port viewed
  by the NBS.  The minimum wavelength is shifted from the
  center of the field of view by tipping the filter by $\approx 2^{\circ}$
  in order to optimize the instrument sensitivity.  Contours and color scale are identical to those in Figure~\ref{fig:ideallambda}.}
\label{fig:rainbowdonut}
\end{figure}

For each pixel in the array, a Gaussian spectral response function was fit to the laser data, yielding measurements of the central wavelength and integral width over the array.
These data and the subsequent fits are shown in Figure \ref{fig:laserfilters} for a subset of individual pixels across the array.  The central wavelength response over the NBS field
is shown in Figure \ref{fig:rainbowdonut}.  The average measured resolution
over all pixels is $\lambda / \Delta \lambda = 1004 \pm 12$ ,
which averages over any spatial variation in wavelength response over the narrow-band filter. 
  The measured system wavelength response is close to the values
listed in Table \ref{tab:optics} based on the measured narrow-band filter, with
a maximum wavelength of $854.51 \pm 0.02 \,$nm.

In addition to the SIRCUS facility, we can verify the reproducibility
of the NBS wavelength response using an integrating sphere coupled to
either a $R \sim 600$ monochromator, a stabilized $852 \,$nm diode
laser and wavemeter, or a Ne lamp with an emission line at $854.5
\,$nm, close to the NBS bandpass.  These were used for quick tests to verify the
wavelength calibration did not change over multiple thermal cycles and after
payload vibration tests.

\subsection{Anomalous Thermal Response}
\label{ssS:thermalresponse}

\begin{figure}[hbtp]
\centering
\includegraphics[width=75mm]{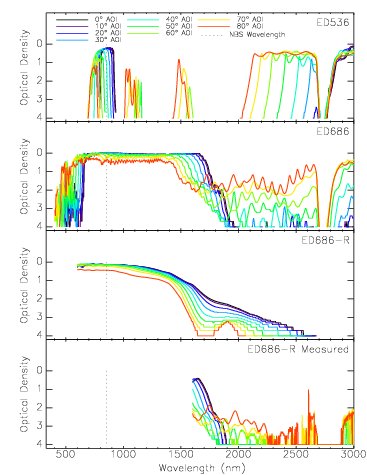}
\caption{Optical Density as a function of incident angle for
  the two blocking filters. ($\mathrm{Optical \, Density} = -\log_{10} (T)$). The long-wavelength blocker ED686 was
  recoated with an absorbing dielectric to improve its large-angle
  blocking.  The panels show from top to bottom 1) calculated
  large-angle transmittance of the ED536 filter; calculated
  performance of the ED686 filter 2) before and 3) after applying the
  absorbing dielectric layer; and 4) the measured large-angle
  transmittance of the ED686 filter with the absorbing coating.  The
  measured transmittance in the 2000 - 2500 range for 60$^{\circ }$ to
  80$^{\circ }$ incident angles appears to not show the full
  calculated improvement and is consistent with the observed reduction
  by factor of $\sim 80$ to thermal emission measured in the
  laboratory.
\label{fig:transmittance}}
\end{figure}

With the original filter set, consisting of the narrow-band filter and
 a single blocking filter (referred to as ED536), installed in the NBS instrument, we
observed a significant background signal of $310 \,$\eps\ when looking
out into a dark laboratory through the fused quartz vacuum window.  By
modulating the temperature of the room-temperature window and holder,
we determined this radiation was thermal emission incident on the
detectors.  The signal greatly exceeded the expected level based on
the measured on-axis filter blocking, though the source of this emission is
off-axis and well outside the field of view.  To reduce this radiation
we added a second blocking filter, ED686 as shown in
Figure~\ref{fig:transmittance}, to improve the long-wavelength attenuation.
However under the same conditions we still observed a large background
level of $110 \,$\eps\ with the new filter in place, certainly an
improvement but not the factor of $> 10^4$ reduction expected from the
additional long-wavelength blocking.

The CIBER instrument was flown in its first flight with the NBS using
this filter configuration.  During the flight the LRS observed a large
long-wavelength signal associated thermal emission from the skin of
the sounding rocket, heated by air friction during ascent
\citep{Tsumura11}.  The skin section had a direct view of the inside
surface of the cold baffle tube and first optical element of all the imagers and spectrometers, and thus
thermal radiation could scatter to the detectors.  Unfortunately the
NBS saw an elevated photo-current of $\sim 150$~e/s$^{-}$ as well,
related to this thermal emission.  After discussions with Barr, we
determined that the filter blocking degrades at large incident angles,
as shown by the model curves in Figure~\ref{fig:transmittance}.  In
order to improve the large-angle performance, Barr designed and fabricated an absorbing
dielectric coating to the long-wavelength ED686 blocking filter, named
ED686-R in Figures \ref{fig:transmittance} and \ref{fig:blocking}.

According to calculations, by adding the absorbing coating, the
large-angle blocking should improve by $10^2$ to $10^4$, over the range
of 0$^{\circ }$ to 80$^{\circ }$ incidence, with little loss of
in-band transmission, from $89.0 \,$\% to $87.5 \,$\%, as measured at
$77 \,$K.  As shown in the bottom panel of
Figure~\ref{fig:transmittance}, the measured large angle performance at
$60^{\circ}$ to $80^{\circ }$ angle of incidence in the $2000 {-} 2500
\,$nm range is $\sim 10^2$ better than the calculated performance
without the coating, but somewhat worse than the calculated
performance with the coating.  The re-coated filter was installed in the instrument,
and we measured a background level that was significantly reduced
looking through the vacuum window, $1.4 \,$\eps, consistent with the
measured improvement of $\sim 80$ at $60^{\circ}$ to $80^{\circ}$.

In addition to improving the filter design, we added a cold
extending/retracting baffle to extend the baffle tube beyond the
heated rocket skin, eliminating all lines of sight from the skin to
the optics and the inside surface of the baffle, described in
detail in \citet{Zemcov11}.  In the second flight the measured thermal
response in the LRS was at least 100 times smaller than in the first
flight, and we did not observe a measurable response in the NBS,
indicating that the baffles and large-angle blocking filters performed
as expected.

\section{Instrumental Systematic Error Control}
\label{S:Laboratory}

\subsection{Requirements}
\label{sS:requirements}

The instrument design controls systematic instrumental errors by
incorporating 1) a cold shutter to measure array dark current; 2) low
scatter optics to eliminate stray light from off-axis sources; 3) a filtering
scheme with excellent spectral blocking to control out-of-band
radiation; 4) a specialized aperture-filling laboratory source to
measure array flat field response; and 5) a calibration program based
on laboratory measurements. 

The laboratory characterization program addresses the systematic error requirements.  In addition to the tests described in Section
\ref{S:testing}, these measurements include sensitivity, filter blocking, dark current,
temperature stability, flat field and calibration determination.


\subsection{Dark Current}
\label{sS:dark}

The generation of charge carriers within the depletion
region of HgCdTe infrared detectors leads to a small ($\sim$0.1 \eps)
dark current even in the absence of incident photons.  In order to
perform absolute measurements of the ZL, this dark
current must be understood and removed from flight data.  In flight,
the dark current is monitored by closing the instrument's cold
shutter prior to launch, during data collection in the middle of the
flight, and again following the vacuum shutter door close event during
reentry.

To determine dark current behavior before flight, we measure the
baseline dark current with the instrument in flight configuration,
with the shielded vacuum door, providing a dark environment in the experiment
section \citep{Zemcov11}.  The NBS cold shutter is closed during the
measurement to ensure the darkest conditions possible at the detector
array.  To determine the dark current, the array is read out for $\sim 1$
hour in this configuration.  Because resetting the array causes a transient
response, the first several frames following a reset are discarded from the
analysis.  We fit a slope and offset over each integration, obtaining an
array median dark current of $0.5 \pm 0.1 \,$\eps, where the uncertainty
is estimated from the variance of twelve 50 s integrations.  The pattern
of the dark current is brightest at the array corners, suggesting residual
emission from the ROIC even though we are not using the final buffer stage.  In the case of the NBS, the dominant dark current characteristic for systematic control is not the mean level, but the stability of spatial morphology across the array.  Gradients or features spuriously aligned with the wavelength response can produce
artificial spectral structures which could mimic the Fraunhofer line.  We quantified this effect through simulations as described in Section~\ref{sS:sensitivity}.

To verify that the dark current is stable over long periods, we
periodically obtained a standard calibration set including a dark
current measurement over the course of several years.  Though
typically taken with the light bulkhead in place of the rocket door,
these measurements show that the dark current is stable over long
periods and many cryogenic cycles.  These long-term dark levels are
consistent with the images obtained on the launch gantry just prior to
flight, and the dark measurements obtained during observations with
the shutter closed.  

We compared dark current measurements taken with a cold lid placed at 77~K over the optics and the cold shutter open and closed, showing a DC offset of 0.33~ $e/s$.  This indicates an additional photocurrent is present, induced by emission from the ROIC reflecting off the NBS optics.  While the amplitude of this signal is large compared to the mean dark current, it has a distinctive stable spatial morphology along the array.  This consists of a spatially flat component and a compact region of excess photocurrent.  The reflected light can be subtracted, or the pixels in the compact region can be simply discarded in the analysis of flight data.  A ZL spectrum was fit to this spatial array response to quantify the systematic contribution of this effect, if no further corrections are applied.  The best fit spectrum had an amplitude of -1.88~\nw, corresponding to a $\sim -0.4\%$ effect for nominal ZL intensities, assuming the calibration
factor determined in Section \ref{S:calibration}.

\subsection{Temperature stability}
\label{sS:stability}

Because the output voltage of the ROIC varies with temperature,
typically $\sim 1000 \, $~e/K, we have implemented a multi-stage
temperature control system.  As described in \citet{Zemcov11}, the
pressure of the cryogenic bath is regulated in flight by an absolute
pressure valve.  In addition, the focal plane temperature is
controlled in 2 thermal stages.  An intermediate stage is actively
regulated.  The focal plane is thermally staged from the controlled
stage to passively filter high-frequency temperature variations.  The
temperature of both stages are precisely monitored using a
high-precision temperature bridge.

To verify that the regulation system works as designed, we cooled the
NBS focal plane to its quiescent temperature (typically $\sim 80 \,$K)
at which point we activated the temperature control unit.  Thermometry
data are collected during the time the focal plane unit takes to reach
a steady operating temperature, which is set in the firmware of the
thermal control unit.  Though the thermal set point is adjustable, it
must be set to a value larger than the highest base temperature so
that it can be controlled in all instrument flight and testing
configurations.  At our chosen set point of $82.23 \,$K, it typically
takes $1.5 \,$h for the heaters to raise the temperature of the focal
plane to its set point $\sim 2 \,$K above the base.  When stabilized,
the active control proportional-integral-derivative (PID) loop
controls the intermediate stage about the set point.  

To quantify the systematic error on ZL intensity measurement induced by thermal instabilities, dark current was measured periodically during the stabilization process.  Images were produced from dark data with thermal drifts ranging up to $1.2$~mK/s. This drift produces a response over the array of typically $\sim1$~e/s.  Because this response is largely spatially uniform, the effect of a thermal drift on ZL signals is minimal.
After subtracting a template of dark current produced by averaging many integrations taken previously with the temperature fully controlled, a Fraunhofer line was fit to the thermally unstable data. The inferred ZL brightness was observed to depend very steeply on thermal instability, with the amplitude of the induced ZL signal falling below the statistical uncertainty (described in Section \ref{sS:sensitivity}) at a drift level of $0.45$~mK/s, equivalent to $\Delta ZL=21.1$~\nw in Table~\ref{tab:performance}.  When fully
controlling, the average drift of the focal plane temperature is $<10
\, \mu$K/s over a 50 s integration, making the effect of thermally varying dark current a negligible systematic.   

The NBS control system is powered on several hours before flight to
ensure that the focal plane temperature has stabilized at launch.
During flight, the temperature of both thermal bridges are telemetered
to the ground.  The flight data show no significant deviation from the
ground behavior during astronomical observations in flight.  Figure 10
of \citet{Zemcov11} shows that the thermal drift of the NBS focal
plane, at its worst during the first hundred seconds of flight, is
$5.3 \, \mu$K/s in the first $50 \,$s of astronomical data
acquisition, which exceeds the specification.  
However, after this period of time the drift is $< 0.02 \, \mu$K/s and thermally-induced drift is negligible.

\subsection{Out of Band Blocking}
\label{sS:Filters}

\begin{figure}[hbtp]
\begin{center}
\includegraphics[width=75mm]{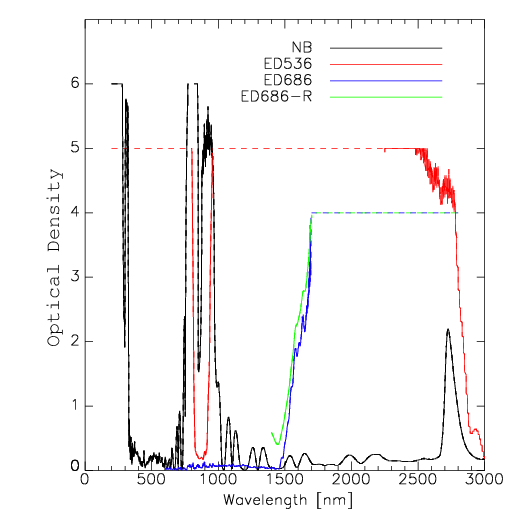}
\end{center}
\caption{Out-of-band blocking of component filters measured at normal
  incidence.  Note
  the straight lines indicate instrumental limits on the out-of-band
  blocking, not actual transmittance.  Of the two blocking filters,
  ED686 was measured before and after an absorbing coating was applied
  to reduce large-angle transmission.}
\label{fig:blocking}
\end{figure}

For precise measurements of the Fraunhofer line brightness, we
require a high degree of out-of-band blocking of the ZL continuum.
The HgCdTe
detectors are  only responsive over a 2100~nm range spanning $400 {-} 2500 \,$nm, which defines the outer wavelength limit for potential contamination. 
The fact that the ZL continuum
peaks at optical wavelengths, and that any spectral leak is unlikely
to be modulated over the field of view as the \ion{Ca}{2} line, further
reduces the systematic error from leaks.  Barr originally designed a
single blocking filter (ED536) to meet the $OD=-\log_{10}(T_{OB}) > 6$
blocking requirement.  This filter (ED536) provides an in-band transmittance of
66 \%.  The blocking performance in Fig.~\ref{fig:blocking} is
excellent, though limited by measurement accuracy to OD$=5$.

The on-axis blocking performance of the NBS was characterized using the tunable long wavelength laser system coupled to the integrating sphere
as described in \citet{Zemcov11}.  The laser wavelength was changed to
discrete values separated by $\sim 100 \,$nm over the range $1100 \leq
\lambda \leq 2300 \,$nm.  At each wavelength step, we determined the
response of the NBS by summing the photocurrent over the array.  To
remove the photocurrent arising from background light in the apparatus,
we recorded ambient frames when the laser was shuttered.  We
differenced the laser on and shuttered data to measure the
photocurrent arising from out-of-band light.

We simultaneously monitored the radiance incident on the NBS using a
calibrated detector viewing the integrating sphere.  Because the laser
is capable of producing a large radiance in a narrow band, the signal
to noise ratios of the measurements are large, even though the
absolute response to out of band light is not.  The average
photocurrent arising from $1100 \leq \lambda \leq 2300$ light is
$2 \times 10^{-5}$ \eps\ / \nw, which is consistent with the OD $> 5$
limit derived from direct filter testing.  We produce an out-of-band
spectral response function 
by interpolating the
measurements over the range $850 \leq \lambda \leq 2500 \,$nm.

To determine if the measured out of band blocking is sufficient, we
estimate the photocurrent expected in flight from ZL by assuming a
Solar ZL spectrum 
for $850 \leq \lambda \leq 2500$.
The expected signal at the NBS detector is then calculated by numerically integrating the product of the ZL spectrum and our out-of-band response function
which yields $i_{\mathrm{OB}} \leq 3 \,$m\eps\ at the ecliptic pole.
This results in a calculated level of $i_{OB}/i_{IB} \leq 0.2 \%$ and is therefore is deemed negligible.  

\subsection{Flat Field Determination}
\label{sS:flatfield}

To measure the Fraunhofer line intensity we must first normalize the
relative responsivity of the individual detector array pixels to a common
value, typically termed the ``flat field'' correction.  Because the
flat field of the NBS detector array cannot be determined independently
of the astrophysical signal in flight, we must measure the flat field
response in the laboratory.  The internal instrument calibration lamp
serves as a transfer standard to verify that the responsivity has not
changed between the laboratory testing and flight.

The flat field measurement consists of a quartz halogen lamp which is
fiber coupled into an integrating sphere inside a vacuum chamber.  We
measured the output spectrum of the integrating sphere and added a filter
at the lamp to synthesize a Solar spectrum over the wavelength range of
interest, $800$~nm$ \leq \lambda \leq 2500$~nm.  The CIBER instrument
shares a common vacuum space with the test chamber so there are no
optical surfaces between the integrating sphere port and the NBS aperture.
Since the illumination pattern of the output of the sphere has been
measured to be uniform to $<$ 0.3\% at angles of incidence $<
10^{\circ}$ the illumination measured at the NBS focal plane is a good
tracer of the array flat field.  

The flat field response is measured by turning on the quartz halogen
lamp, recording an image, and then subtracting images before and after
the lamp is off to remove any background signal.  Finally these measurements are bracketed by dark
frames to ensure that the array baseline behavior is stable. The flat
field is measured at a variety of input flux levels from 1000 to 10 \eps\
to ensure the flat field measurement does not vary significantly with
source brightness due to detector non-linearity.

To quantify the stability of the flat field and its effect on the instrument's ability to accurately measure the depth of the Fraunhofer line, the measurement was repeated three independent times.  Each measurement was conducted in a separate cooldown of the cryostat. To quantify stability, the amplitude of the Fraunhofer line was fit in a direct measurement of the Solar spectrum (described in detail in Section \ref{S:solmeas}) using each of the three independent flat field measurements.  The measured amplitude was shown to vary at the 1.4$\%$ level between each of the three measurements.  

\subsection{Stray Light Performance}
\label{sS:Straylight}

The NBS optics must also reject sources of off-axis emission.  In the case of CIBER, this is dominated by stray light from the Earth.  The response to extended off-axis
emission is given by \citet{Bock95} to be

\begin{equation}
\label{eq:straylight}
I_{stray} = \frac{1}{4 \pi} \int g(\theta) I_{s}(\theta,\phi) d\Omega, 
\end{equation}
where $I_{stray}$ is the stray light signal observed surface brightness on
the sky, $I_{s}$ is the surface brightness of the emitting off-axis source, and
$g(\theta) = \frac{4 \pi}{\Omega_{FOV}} G(\theta)$ is the telescope gain function, where
$G(\theta)$ is the telescope response to an off-axis point source normalized to
unity on axis, and $\Omega_{FOV}$ is the telescope field of view.

To measure $G(\theta)$ in the laboratory, we followed the procedure described in \citet{Tsumura11}.  This consisted of replacing the PICNIC array with a warm silicon photo-diode\footnote{S10043 manufactured by Hamamatsu photonics (http://www.hamamatsu.com)} and mounting the NBS to a custom optical bench capable of rotation to 90$^\circ$ from on-axis.  A collimated halogen source was chopped at 20~Hz as the angle to the NBS boresight was varied. The amplitude of the modulated signal was measured with a lock-in amplifier referenced to the signal from the chopper.  The dynamic range of this measurement was limited due to the small bandwidth of the NBS such that the largest angle with sufficient signal to noise was 25$^\circ$ from center.  At this angle, $G (25^\circ) = 5 \times 10^{-5}$ and the function displayed a sharp slope, suggesting it continues to fall at higher angles.  
 
 The night-side brightness of the earth in the NBS band is dominated
by airglow emission and is approximately $10^{3}$ nW m$^{-2}$ sr$^{-1}$
\citep{Leinert98} 
at 854 nm.  As the earth limb is $> 55^{\circ}$ off-axis during observations,
the maximum projected solid angle is $\Omega_{\oplus} = 0.2 \,$ sr.
Even in the overly conservative case in which $G(\theta)$ is assumed to remain flat at $5 \times 10^{-5}$ to higher angles than were measured in the lab, the integral in Equation~\ref{eq:straylight} results in a signal $I_{stray} < 1$~nW m$^{-2}$ sr$^{-1}$.  We therefore conclude that off-axis emission from the earth contributes a negligible systematic error to the measurement.

\subsection{Calibration}
\label{S:calibration}

The NBS must be accurately calibrated in order to compare our ZL
measurements with ZL models based on DIRBE observations.  Specifically,
we need to translate the NBS ZL measurements to the DIRBE $1.25 \, \mu$m
band with an accuracy of $\lesssim 5 \,$\% to distinguish between models.
To perform this comparison, we must absolutely calibrate the NBS, then
translate the ZL brightness measured at $854 \,$nm to $1.25 \, \mu$m using
the CIBER-LRS ZL spectral color.  To reduce the contribution of the uncertainty
in the NBS calibration to the overall uncertainty in this chain, the NBS must be accurately calibrated relative to DIRBE.  For comparison,
 the absolute calibration uncertainty in standard calibration
stars like Vega is typically $\sim~1\%$ and the $1.25~\mu$m DIRBE channel absolute
calibration uncertainty is 3\% \citep{Burdick1997} obtained on Sirius.  In addition to the laboratory calibration we describe below, the NBS observes Vega in every flight for an independent calibration check.

A schematic representation of the NBS calibration set up is shown in
Figure 15 of \citet{Zemcov11}, where light from a stable source is coupled
into an integrating sphere which is viewed by the NBS and a reference detector
absolutely calibrated by NIST.  The radiance of the light source is varied
and the response of the NBS is referenced to the detector.  

Our primary photometric calibration measurements are derived from the monochromatic data described in detail in Section~\ref{sS:wavelength} which use the SIRCUS laser facility as the stabilized light source \citep{Brown06}.  
The reference detectors which measure absolute radiance are constructed of single element large area Si photodiodes and baffles that limit the angular extent light reaching the detector. A smaller sphere feeding the larger sphere, which is viewed by the NBS,  is constantly monitored with Si photodiodes to ensure the light levels remain known during the NBS calibration given the ratio of large sphere radiance to monitor signal previously determined with the reference detectors. Uncertainties in the deduced narrow band radiance are approximately 0.3\%, dominated by contributions from the sphere uniformity and the effects of out-of-field light.

Photocurrent measurements using the laser source determine the NBS absolute responsivity,  at a discrete set of wavelengths. 
We calculate an array-wide map of the calibration factor 
($ \frac{\int I_{\lambda^{\prime}} d \lambda^{\prime}}  {i}   $)
for a spectrally flat source from the ratio of the measured NBS photocurrent and absolute radiance. 
The integral is evaluated from scans of the SIRCUS laser across a 100~nm wide range.
After applying the flat field correction, described in Section~\ref{sS:flatfield}, the array wide mean response factor is ($602  \pm 10 \,$\nw)~/~(e$^{-}$ s$^{-1}$) after correcting for the transmittance of the additional window used in the laboratory. The $1\sigma$ uncertainty in the quoted absolute calibration factor is determined by the variance across all pixels in the array, added in quadrature with the $0.3\%$ systematic uncertainty from the absolute calibration quoted by NIST. 

This calibration factor was checked by coupling a supercontiuum laser source to the integrating sphere referenced to a detector system capable of $R
\sim 500$ spectral measurements, calibrated to measure radiance traceable to fixed-point blackbodies.  The supercontinuum laser produces a relatively flat featureless spectrum across the NBS band.  
Measurements of the photocurrent induced by the broad band source were taken at four light levels ranging from 500 to 2000~e$^{-} /s$ while simultaneously monitoring the absolute radiance of the sphere.  Dark signals were referenced to the ambient light background of the laboratory.  Each exposure is corrected using the flat field template described in Section~\ref{sS:flatfield}. The absolute calibration factor is taken to be the slope of the linear correlation between the NIST measured absolute radiance and the mean photocurrent of each exposure.  The resulting value is ($603  \pm 12 \,$\nw)~/~(e$^{-}$ s$^{-1}$), after correcting for the transmittance of the additional window used in the laboratory, in excellent agreement with the measurement from the narrow band source.  The quoted uncertainty is dominated by instrument noise in the NIST spectrographs and linearity during the boot-strapping measurements which pushes the dynamic range of the reference spectrographs.  

The SIRCUS data set provides absolute calibration and a full spectral response measurement for every pixel.  This allows for a more sophisticated treatment of flight data.  The response over the array may be calculated for any source spectrum using Equation~\ref{eq:Ilambda}.  Thus we can make template response maps over the array to ZL and EBL that fully include flat field gain and variation in the spectral response over the array.

\subsection{Sensitivity}
\label{sS:sensitivity}

We estimate the complete instrument sensitivity by simulating a series of $50 \,$s observations using measured dark data as input noise.
The data used in this calculation were taken in full flight configuration on the launch gantry one day before the second CIBER flight with the cold shutter closed.  There were 30 independent $50$~s integrations obtained.  The variance between instances contains fluctuations in both dark current spatial morphology and correlated read noise.
We added a model ZL spectrum convolved with the instrument response (e.g. Figure~\ref{fig:rainbowdonut}) to each independent integration.  The amplitude of the Fraunhofer line in the simulated observation was then fit and compared to the input value.  The measured 50~s integration sensitivity is taken to be the standard deviation of the distribution of residuals (input sky brightness - measured sky brightness), which includes all sources of correlated and uncorrelated noise over the array.  The sensitivity is determined to be $\Delta ZL = 21.1 \,$ nW m$^{-2}$ sr$^{-1}$, shown in Table
\ref{tab:performance}, converted to ZL units assuming the calibration
factor from Section \ref{S:calibration}.
Since the flight photocurrent
varies between $0.5$ and $1.5 \,$\eps, the contribution of photon noise
is small.  Given the typical brightness of $500 \,$\nw\ for high ecliptic
latitude ZL brightness \citep{Tsumura10}, this leads to a $\Delta ZL /
ZL \lesssim 4.4 \,$\% accuracy on the ZL brightness in a 50~s observation.  

\begin{table*}[htb]
\centering
\caption{Measured System-Level Instrument Characteristics.}
\begin{tabular}{lc}
\hline
Parameter & Measured Value \\ \hline

Mean Dark Current Stability & $0.5 \pm 0.1 \,$\eps \\

Focal Plane Temperature Stability & $<5.3 \, \mu$K/s \\

Spectral Resolution $\lambda / \Delta \lambda$ & 1004 $\pm$ 12  \\

Line Contrast & $26 \,$\% \\

Lab Responsivity & {($602  \pm 10 \,$\nw) / (e$^{-}$ s$^{-1}$)} \\ 

Out of band spectral response & $T_{OB} < 10^{-6}$ for $0.4 \leq \lambda \leq 2.5\, \mu$m \\

Flat field response & $\Delta FF / FF$ gives $\Delta ZL/ZL = 1.4$ \% \\

ZL sensitivity in $50 \,$s ($1 \sigma$) &  {$\Delta ZL=21.1 \,$\nw} \\

\end{tabular}
\label{tab:performance}
\end{table*}

\section{Laboratory Measurement of the Solar Spectrum}
\label{S:solmeas}

As the NBS is designed to measure absorption in Solar light scattered off the interplanetary dust cloud, a high signal to noise measurement of sunlight is a useful tool for testing end to end system functionality.
To obtain these data, we pointed a long optical fiber at the Sun and coupled the other end to the integrating sphere to uniformly illuminate the NBS aperture.  The coupling efficiency to the sphere was optimized to produce an array photocurrent near 200~e/s, where the signal to noise is very high in each pixel without nearing saturation.  We measured the Solar spectrum presented from a 200~s exposure
and the flat-field correction described in Section \ref{sS:flatfield}.  

\begin{figure}[hbtp]
\includegraphics[width=75mm]{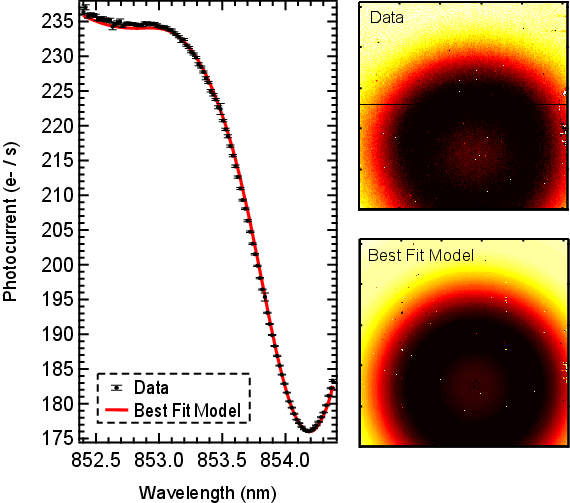}
\caption{{\it Left}: Measured and best fit 1D Solar spectra obtained from coupling sunlight to the integrating sphere in the laboratory. The calculated Solar response was generated from the high resolution measurements of \citet{Debouille1990} convolved with the measured NBS spectral response. Error
  bars are 1 $\sigma$ statistical.  {\it Top right}: Flat field
  corrected 2D measured Solar spectrum. {\it Bottom right}: Calculated Solar response with fitted amplitude and offset.  The color scales are matched in the 2D images.}
\label{fig:meassolar}
\end{figure}

For comparison, we calculated a predicted model spectrum by convolving the high resolution Solar measurements of \citet{Debouille1990} with the instrument spectral response function obtained from discrete-wavelength SIRCUS laser data.
The model was fit to the data with a simple $\chi^2$ minimization using two free parameters: a normalization factor and an additive offset to account for stray light from the laboratory. The measured Solar spectrum is shown in Figure \ref{fig:meassolar} in both one and two dimensions alongside the best fit model.  With only two free parameters, the maximum deviation of the residual
of data from model is 0.24$\%$.

\section{Second Flight Performance}
\label{S:second flight}

The CIBER instrument was flown on 11 July 2010 from White Sands
Missile Range.  During this flight the NBS performed well, with the
shutter, calibration lamp, extendable/retractable baffle and array all
operating normally.  All flight events occurred according to the
planned sequence, and the attitude control system gave stable
pointing on all of the astronomical fields.  The NBS obtained images
of the sky in flight consistent with focus data
from laboratory measurements.  The CIBER instrument was recovered for
post-flight testing and future flights. 

The NBS photocurrents in the second flight range between $2
\,$\eps\ at an ecliptic latitude of $10^{\circ}$ to $1 \,$\eps\ at an
ecliptic latitude of $90^{\circ}$.  Based on the laboratory
calibration, these levels correspond to a ZL brightness ranging from
500 to 1000 nW m$^{-2}$ sr$^{-1}$, which matches a simple
extrapolation of the DIRBE sky brightness to NBS wavelengths.  There is no measurable thermal response in the NBS second
flight data.  The in-flight dark level obtained with the shutter
closed was 0.43 e$^{-}$/s, consistent with the pre-flight dark images
obtained on the launch gantry.  The array temperature was stable in
flight, giving a maximum drift of 5.3 $\mu$ K / s over a 50 s
integration during astronomical observations.  

\section{Astrophysical Systematics}
\label{S:astrosys}

In addition to instrumental effects, estimating the ZL intensity from a Fraunhofer line is subject to several astrophysical systematic errors.  

\subsection{Dynamical Effects}
\label{sS:dynam}

The Fraunhofer line profile can be altered by dynamical effects within the interplanetary dust cloud. \citet{wham} have done an extensive study of these effects using the \ion{Mg}{1} Fraunhofer line at 581.4~nm along various lines of sight.  The Doppler inferred velocity of the line was found to depend strongly on the Solar elongation angle, $\epsilon$, with peak magnitudes of $\sim$15~km/s around $\epsilon = 90^{\circ}$.  At the spectral resolution  of the NBS (shown in Table~\ref{tab:performance}), this effect is barely detectable ($\sim10\times$ smaller than the instrument FWHM).  To quantify uncertainty caused by velocity variation, we simulated shifting the ZL Fraunhofer line across a conservative range spanning $\pm$ 20~km/s from rest and fitting the resulting spectrum assuming zero.  The best fit ZL amplitude was altered by at most 0.3\%. 

\citet{wham} also find a kinematic broadening of the Fraunhofer line corresponding to a velocity dispersion of $\sim15$~km/s.  The \ion{Ca}{2} line has an intrinsic width of 70~km/s and the FWHM of the NBS bandpass is 245~km/s. Since the line is not fully resolved by the NBS and the effects of broadening are small compared to the intrinsic width, the dynamical effects have negligible impact on total ZL brightness measurements.

\subsection{Galactic Fraunhofer Line Foregrounds}
\label{sS:caII}

NIR EBL measurements are subject to Galactic foregrounds consisting of Integrated Stellar Light (ISL) from unmasked stars and Diffuse Galactic Light (DGL), caused by starlight scattered by interstellar dust.  Both of these signals will contain Fraunhofer lines at some level.  Detected stars are masked and removed from analysis, but the aggregate signal from fainter stars is present and must be modeled.  

\citet{Bernstein2002II} have generated ISL spectral templates at $0.4$~nm resolution in the optical which contain the \ion{Ca}{2} H (396.85~nm) and K (393.37~nm) lines. Both lines appear shallower in the ISL than in the solar spectrum, but by different amounts.  For this reason, a generalized conclusion about the depth of Fraunhofer lines in the ISL relative to Solar would be  inaccurate.  Unfortunately, these high resolution spectra do not cover the \ion{Ca}{2} triplet and are therefore not directly applicable to this study.  Instead, for this estimate we apply the lower resolution ISL template of 
\citet{Mattila80I,Mattila1980}, which spans the UV to NIR at $5$~nm resolution. This resolution is insufficient to individually resolve all of the \ion{Ca}{2} triplets in the band.  By comparing high-resolution measurements of the Solar spectrum \citep{Debouille1990} convolved with a $5$~nm Gaussian response function, we estimate that the \ion{Ca}{2} triplet appears at near Solar EW in the ISL.

Fraunhofer lines in DGL have also been detected in the optical \citep{Mattila2011}. As DGL arises from stellar light scattered by diffuse dust clouds, we assume the EW of the Fraunhofer line in the DGL is the same as it is in the ISL.  The total contribution from DGL can be modeled by scaling the measured DGL to 100~$\mu$m dust emission \citep{Matsuoka2011} as well as neutral hydrogen column density \citep{Mattila2011}.

To estimate the expected level of the Galactic Fraunhofer signal, we take the example of a field in the direction of Bootes observed by CIBER (ecliptic longitude of 200.57$^{\circ}$ and latitude of	46.633$^{\circ}$) and assume that both DGL and ISL contribute \ion{Ca}{2} absorption features with Solar EW.  The level of the ISL is determined from the star counts predicted by the Trilegal model code \citep{trilegal} below a cutoff magnitude of 14, comparable to the expected completeness in an NBS image. The DGL intensity is estimated using the correlation with neutral hydrogen column density ($n_{H}$) proposed by \citet{Mattila2011}, with $n_{H} = 1.03 \times 10^{20}$~cm$^{-2}$ \citep{DL1990}. The ZL intensity is taken from \citet{Kelsall98}, scaled to NBS wavelengths assuming the spectral template generated from the CIBER-LRS \citep{Tsumura10}.  With the aforementioned assumptions, we estimate a contribution at $854$~nm of 15~\nw from ISL, 10~\nw from DGL and 386~\nw from ZL, or a 6.5\%  contribution from the Galactic Fraunhofer line signal to the total sky brightness.  This Galactic contribution can be assessed and removed with an improved ISL model spectrum.

\subsection{Raman Scattering}
\label{sS:raman}
If significant Raman scattering effects are present in the ZL, Fraunhofer lines will be filled, causing the relationship between line depth and continuum brightness to be systematically biased.  \citet{Dube79} compare this ratio for a Fraunhofer line in sunlight reflected off the moon to the ZL, and find them to agree within 1.5\%, limited by instrumental error.  This suggests that if present at all, Raman scattering is a small effect, consistent with the notion that
 interplanetary dust particles have large dielectric constants with few free electrons.

\section{Summary}
\label{S:summary}

We describe the design and performance of a specialized wide-field,
narrow-band spectrometer for measuring the absolute ZL brightness
using the EW of the 854.2 nm \ion{Ca}{2} Fraunhofer line,
viewed in sunlight scattered by interplanetary dust.  
The systematics limiting the uncertainty, both instrumental and astrophysical, are discussed and quantified through an extensive suite of laboratory characterization and simulation.  The aggregate instrumental uncertainty is dominated by read noise fluctuations and flat field error.

The instrument was flown on a second sounding rocket flight in July
2010 and demonstrated expected performance, with nominal photo
currents, temperature stability, calibration lamp response, and dark
currents.  Preliminary reductions of the flight data show clear significant detections of the Fraunhofer line. A third flight was carried out successfully in March 2012.  In-depth analysis of the science data is currently underway and results will be released in the near future.  

\section*{Acknowledgements}

This work was supported by NASA APRA research grants NNX07AI54G,
NNG05WC18G, NNX07AG43G, NNX07AJ24G, and NNX10AE12G.  Initial support
was provided by an award to J.B.~from the Jet Propulsion Laboratory's
Director's Research and Development Fund.  Japanese participation in
CIBER was supported by KAKENHI (20$\cdot$34, 18204018, 19540250,
21340047 and 21111004) from Japan Society for the Promotion of Science
(JSPS) and the Ministry of Education, Culture, Sports, Science and
Technology (MEXT).  Korean participation in CIBER was supported by the
Pioneer Project from Korea Astronomy and Space science Institute
(KASI).

Certain commercial equipment, instruments, or materials are identified
in this paper to foster understanding. Such identification does not
imply recommendation or endorsement by the National Institute of
Standards and Technology, nor does it imply that the materials or
equipment identified are necessarily the best available for the
purpose.

This work utilizes SOLIS data obtained by the NSO Integrated Synoptic 
Program (NISP), managed by the National Solar Observatory, which is 
operated by the Association of Universities for Research in Astronomy (AURA), 
Inc. under a cooperative agreement with the National Science Foundation.

We would like to acknowledge the dedicated efforts of the sounding
rocket staff at the NASA Wallops Flight Facility and the White Sands
Missile Range.  We also acknowledge the work of the Genesia
Corporation for technical support of the CIBER optics.
A.C.~acknowledges support from an NSF CAREER award, B.K.~acknowledges
support from a UCSD Hellman Faculty Fellowship, K.T.~acknowledges
support from the JSPS Research Fellowship for Young Scientists, and
M.Z. and P.M.K~acknowledge support from NASA Postdoctoral Fellowship.


\begin{thebibliography}{55}
\expandafter\ifx\csname natexlab\endcsname\relax\def\natexlab#1{#1}\fi

\bibitem[{{Aharonian} {et~al.}(2006){Aharonian}, {Akhperjanian}, {Bazer-Bachi},
  {Beilicke}, {Benbow}, {Berge}, {Bernl{\"o}hr}, {Boisson}, {Bolz}, {Borrel},
  {Braun}, {Breitling}, {Brown}, {Chadwick}, {Chounet}, {Cornils},
  {Costamante}, {Degrange}, {Dickinson}, {Djannati-Ata{\"i}}, {Drury}, {Dubus},
  {Emmanoulopoulos}, {Espigat}, {Feinstein}, {Fontaine}, {Fuchs}, {Funk},
  {Gallant}, {Giebels}, {Gillessen}, {Glicenstein}, {Goret}, {Hadjichristidis},
  {Hauser}, {Hauser}, {Heinzelmann}, {Henri}, {Hermann}, {Hinton}, {Hofmann},
  {Holleran}, {Horns}, {Jacholkowska}, {de Jager}, {Kh{\'e}lifi}, {Klages},
  {Komin}, {Konopelko}, {Latham}, {Le Gallou}, {Lemi{\`e}re},
  {Lemoine-Goumard}, {Leroy}, {Lohse}, {Martin}, {Martineau-Huynh},
  {Marcowith}, {Masterson}, {McComb}, {de Naurois}, {Nolan}, {Noutsos},
  {Orford}, {Osborne}, {Ouchrif}, {Panter}, {Pelletier}, {Pita},
  {P{\"u}hlhofer}, {Punch}, {Raubenheimer}, {Raue}, {Raux}, {Rayner}, {Reimer},
  {Reimer}, {Ripken}, {Rob}, {Rolland}, {Rowell}, {Sahakian}, {Saug{\'e}},
  {Schlenker}, {Schlickeiser}, {Schuster}, {Schwanke}, {Siewert}, {Sol},
  {Spangler}, {Steenkamp}, {Stegmann}, {Tavernet}, {Terrier}, {Th{\'e}oret},
  {Tluczykont}, {van Eldik}, {Vasileiadis}, {Venter}, {Vincent}, {V{\"o}lk}, \&
  {Wagner}}]{Aharonian2006}
{Aharonian}, F., {et~al.} 2006, \nat, 440, 1018

\bibitem[{{Allen}(1976)}]{Allen1976}
{Allen}, C.~W. 1976, {Astrophysical Quantities}, ed. {Allen, C.~W.}

\bibitem[{{Beggs} {et~al.}(1964){Beggs}, {Blackwell}, {Dewhirst}, \&
  {Wolstencroft}}]{Beggs1964}
{Beggs}, D.~W., {Blackwell}, D.~E., {Dewhirst}, D.~W., \& {Wolstencroft}, R.~D.
  1964, \mnras, 127, 329

\bibitem[{{Bernstein} {et~al.}(2002{\natexlab{a}}){Bernstein}, {Freedman}, \&
  {Madore}}]{BernsteinZL}
{Bernstein}, R.~A., {Freedman}, W.~L., \& {Madore}, B.~F. 2002{\natexlab{a}},
  \apj, 571, 85

\bibitem[{{Bernstein} {et~al.}(2002{\natexlab{b}}){Bernstein}, {Freedman}, \&
  {Madore}}]{Bernstein2002II}
---. 2002{\natexlab{b}}, \apj, 571, 85

\bibitem[{{Bernstein} {et~al.}(2002{\natexlab{c}}){Bernstein}, {Freedman}, \&
  {Madore}}]{Bernstein2002III}
---. 2002{\natexlab{c}}, \apj, 571, 107

\bibitem[{{Bernstein} {et~al.}(2005){Bernstein}, {Freedman}, \&
  {Madore}}]{Bernstein2005_corr}
---. 2005, \apj, 632, 713

\bibitem[{{Berta} {et~al.}(2010){Berta}, {Magnelli}, {Lutz}, {Altieri},
  {Aussel}, {Andreani}, {Bauer}, {Bongiovanni}, {Cava}, {Cepa}, {Cimatti},
  {Daddi}, {Dominguez}, {Elbaz}, {Feuchtgruber}, {F{\"o}rster Schreiber},
  {Genzel}, {Gruppioni}, {Katterloher}, {Magdis}, {Maiolino}, {Nordon},
  {P{\'e}rez Garc{\'{\i}}a}, {Poglitsch}, {Popesso}, {Pozzi}, {Riguccini},
  {Rodighiero}, {Saintonge}, {Santini}, {Sanchez-Portal}, {Shao}, {Sturm},
  {Tacconi}, {Valtchanov}, {Wetzstein}, \& {Wieprecht}}]{Berta2010}
{Berta}, S., {et~al.} 2010, \aap, 518, L30+

\bibitem[{{Bock} {et~al.}(2012){Bock}, {Sullivan}, {Arai}, {Battle}, {Cooray},
  {Hristov}, {Keating}, {Kim}, {Lam}, {Lee}, {Levenson}, {Mason}, {Matsumoto},
  {Matsuura}, {Mitchell-Wynne}, {Nam}, {Renbarger}, {Smidt}, {Suzuki},
  {Tsumura}, {Wada}, \& {Zemcov}}]{Bock2012}
{Bock}, J., {et~al.} 2012, ArXiv e-prints

\bibitem[{{Bock} {et~al.}(1994){Bock}, {Lange}, {Matsumoto}, {Eisenhardt},
  {Hacking}, \& {Schember}}]{Bock1994}
{Bock}, J.~J., {Lange}, A.~E., {Matsumoto}, T., {Eisenhardt}, P.~B., {Hacking},
  P.~B., \& {Schember}, H.~R. 1994, Experimental Astronomy, 3, 119

\bibitem[{{Bock} {et~al.}(1995){Bock}, {Lange}, {Onaka}, {Matsuhara},
  {Matsumoto}, \& {Sato}}]{Bock95}
{Bock}, J.~J., {Lange}, A.~E., {Onaka}, T., {Matsuhara}, H., {Matsumoto}, T.,
  \& {Sato}, S. 1995, \ao, 34, 2268

\bibitem[{{Brown} {et~al.}(2006){Brown}, {Eppeldauer}, \& {Lykke}}]{Brown06}
{Brown}, S.~W., {Eppeldauer}, G.~P., \& {Lykke}, K.~R. 2006, \ao, 45, 8218

\bibitem[{{Burdick} \& {Murdock}(1997)}]{Burdick1997}
{Burdick}, S.~V., \& {Murdock}, T.~L. 1997, {COBE Final Report: DIRBE Celestial
  Calibration}, Tech. rep.

\bibitem[{{Cambr{\'e}sy} {et~al.}(2001){Cambr{\'e}sy}, {Reach}, {Beichman}, \&
  {Jarrett}}]{Cambresy2001}
{Cambr{\'e}sy}, L., {Reach}, W.~T., {Beichman}, C.~A., \& {Jarrett}, T.~H.
  2001, \apj, 555, 563

\bibitem[{{Delbouille} {et~al.}(1990){Delbouille}, {Roland}, \&
  {Neven}}]{Debouille1990}
{Delbouille}, L., {Roland}, G., \& {Neven}, L. 1990, {Atlas photometrique DU
  spectre solaire de [lambda] 3000 a [lambda] 10000}, ed. {Delbouille, L.,
  Roland, G., \& Neven, L.}

\bibitem[{{Dickey} \& {Lockman}(1990)}]{DL1990}
{Dickey}, J.~M., \& {Lockman}, F.~J. 1990, \araa, 28, 215

\bibitem[{{Dube} {et~al.}(1977){Dube}, {Wickes}, \& {Wilkinson}}]{Dube77}
{Dube}, R.~R., {Wickes}, W.~C., \& {Wilkinson}, D.~T. 1977, \apjl, 215, L51

\bibitem[{{Dube} {et~al.}(1979){Dube}, {Wickes}, \& {Wilkinson}}]{Dube79}
---. 1979, \apj, 232, 333

\bibitem[{{Dwek} \& {Arendt}(1998)}]{Dwek1998}
{Dwek}, E., \& {Arendt}, R.~G. 1998, \apjl, 508, L9

\bibitem[{{Dwek} {et~al.}(2005){Dwek}, {Arendt}, \& {Krennrich}}]{Dwek05}
{Dwek}, E., {Arendt}, R.~G., \& {Krennrich}, F. 2005, \apj, 635, 784

\bibitem[{{Eather} \& {Reasoner}(1969)}]{Eather1969}
{Eather}, R.~H., \& {Reasoner}, D.~L. 1969, \ao, 8, 227

\bibitem[{{Fixsen} {et~al.}(1998){Fixsen}, {Dwek}, {Mather}, {Bennett}, \&
  {Shafer}}]{Fixsen1998}
{Fixsen}, D.~J., {Dwek}, E., {Mather}, J.~C., {Bennett}, C.~L., \& {Shafer},
  R.~A. 1998, \apj, 508, 123

\bibitem[{{Hauser} {et~al.}(1998){Hauser}, {Arendt}, {Kelsall}, {Dwek},
  {Odegard}, {Weiland}, {Freudenreich}, {Reach}, {Silverberg}, {Moseley},
  {Pei}, {Lubin}, {Mather}, {Shafer}, {Smoot}, {Weiss}, {Wilkinson}, \&
  {Wright}}]{Hauser1998}
{Hauser}, M.~G., {et~al.} 1998, \apj, 508, 25

\bibitem[{{Hodapp} {et~al.}(1996){Hodapp}, {Hora}, {Hall}, {Cowie}, {Metzger},
  {Irwin}, {Vural}, {Kozlowski}, {Cabelli}, {Chen}, {Cooper}, {Bostrup},
  {Bailey}, \& {Kleinhans}}]{Hodapp1996}
{Hodapp}, K., {et~al.} 1996, New Astronomy, 1, 177

\bibitem[{{Jacquinot}(1954)}]{Jacquinot1954}
{Jacquinot}, P. 1954, Journal of the Optical Society of America (1917-1983),
  44, 761

\bibitem[{{Juvela} {et~al.}(2009){Juvela}, {Mattila}, {Lemke}, {Klaas},
  {Leinert}, \& {Kiss}}]{Juvela2009}
{Juvela}, M., {Mattila}, K., {Lemke}, D., {Klaas}, U., {Leinert}, C., \&
  {Kiss}, C. 2009, \aap, 500, 763

\bibitem[{{Keenan} {et~al.}(2010){Keenan}, {Barger}, {Cowie}, \&
  {Wang}}]{Keenan2010}
{Keenan}, R.~C., {Barger}, A.~J., {Cowie}, L.~L., \& {Wang}, W.-H. 2010, \apj,
  723, 40

\bibitem[{{Kelsall} {et~al.}(1998){Kelsall}, {Weiland}, {Franz}, {Reach},
  {Arendt}, {Dwek}, {Freudenreich}, {Hauser}, {Moseley}, {Odegard},
  {Silverberg}, \& {Wright}}]{Kelsall98}
{Kelsall}, T., {et~al.} 1998, \apj, 508, 44

\bibitem[{{Kutyrev} {et~al.}(2008){Kutyrev}, {Arendt}, {Dwek}, {Moseley},
  {Rapchun}, \& {Silverberg}}]{Kutyrev2008}
{Kutyrev}, A.~S., {Arendt}, R., {Dwek}, E., {Moseley}, S.~H., {Rapchun}, D., \&
  {Silverberg}, R.~F. 2008, in Presented at the Society of Photo-Optical
  Instrumentation Engineers (SPIE) Conference, Vol. 7014, Society of
  Photo-Optical Instrumentation Engineers (SPIE) Conference Series

\bibitem[{{Lee} {et~al.}(2010){Lee}, {Kim}, {Tsumura}, {Zemcov}, {Nam}, {Bock},
  {Battle}, {Hristov}, {Renbarger}, {Matsumoto}, {Sullivan}, {Levenson},
  {Mason}, {Matsuura}, \& {Kim}}]{Lee2010}
{Lee}, D.~H., {et~al.} 2010, Journal of Astronomy and Space Sciences, 27, 401

\bibitem[{{Leinert} {et~al.}(1998){Leinert}, {Bowyer}, {Haikala}, {Hanner},
  {Hauser}, {Levasseur-Regourd}, {Mann}, {Mattila}, {Reach}, {Schlosser},
  {Staude}, {Toller}, {Weiland}, {Weinberg}, \& {Witt}}]{Leinert98}
{Leinert}, C., {et~al.} 1998, \aaps, 127, 1

\bibitem[{{Levenson} {et~al.}(2007){Levenson}, {Wright}, \&
  {Johnson}}]{Levenson2007}
{Levenson}, L.~R., {Wright}, E.~L., \& {Johnson}, B.~D. 2007, \apj, 666, 34

\bibitem[{{Madau} \& {Silk}(2005)}]{Madau2005}
{Madau}, P., \& {Silk}, J. 2005, \mnras, 359, L37

\bibitem[{{Marsden} {et~al.}(2009){Marsden}, {Ade}, {Bock}, {Chapin}, {Devlin},
  {Dicker}, {Griffin}, {Gundersen}, {Halpern}, {Hargrave}, {Hughes}, {Klein},
  {Mauskopf}, {Magnelli}, {Moncelsi}, {Netterfield}, {Ngo}, {Olmi}, {Pascale},
  {Patanchon}, {Rex}, {Scott}, {Semisch}, {Thomas}, {Truch}, {Tucker},
  {Tucker}, {Viero}, \& {Wiebe}}]{Marsden2009}
{Marsden}, G., {et~al.} 2009, \apj, 707, 1729

\bibitem[{{Martin}(1988)}]{Martin1988}
{Martin}, P.~G. 1988, \apjs, 66, 125

\bibitem[{{Matsumoto} {et~al.}(2005){Matsumoto}, {Matsuura}, {Murakami},
  {Tanaka}, {Freund}, {Lim}, {Cohen}, {Kawada}, \& {Noda}}]{Matsumoto05}
{Matsumoto}, T., {et~al.} 2005, \apj, 626, 31

\bibitem[{{Matsuoka} {et~al.}(2011){Matsuoka}, {Ienaka}, {Kawara}, \&
  {Oyabu}}]{Matsuoka2011}
{Matsuoka}, Y., {Ienaka}, N., {Kawara}, K., \& {Oyabu}, S. 2011, \apj, 736, 119

\bibitem[{{Matsuura} {et~al.}(2011){Matsuura}, {Shirahata}, {Kawada},
  {Takeuchi}, {Burgarella}, {Clements}, {Jeong}, {Hanami}, {Khan}, {Matsuhara},
  {Nakagawa}, {Oyabu}, {Pearson}, {Pollo}, {Serjeant}, {Takagi}, \&
  {White}}]{Matsuura2011}
{Matsuura}, S., {et~al.} 2011, \apj, 737, 2

\bibitem[{{Mattila}(1980{\natexlab{a}})}]{Mattila80I}
{Mattila}, K. 1980{\natexlab{a}}, \aaps, 39, 53

\bibitem[{{Mattila}(1980{\natexlab{b}})}]{Mattila1980}
---. 1980{\natexlab{b}}, \aap, 82, 373

\bibitem[{{Mattila}(1980{\natexlab{c}})}]{Mattila80II}
---. 1980{\natexlab{c}}, \aap, 82, 373

\bibitem[{{Mattila}(2003)}]{Mattila2003}
---. 2003, \apj, 591, 119

\bibitem[{{Mattila} {et~al.}(2011){Mattila}, {Lehtinen}, {Vaisanen}, {von
  Appen-Schnur}, \& {Leinert}}]{Mattila2011}
{Mattila}, K., {Lehtinen}, K., {Vaisanen}, P., {von Appen-Schnur}, G., \&
  {Leinert}, C. 2011, ArXiv e-prints

\bibitem[{{Misconi}(1976)}]{Misconi76}
{Misconi}, N.~Y. 1976, \aap, 51, 357

\bibitem[{{P{\'e}nin} {et~al.}(2011){P{\'e}nin}, {Lagache}, {Noriega-Crepo},
  {Grain}, {Miville-Desch{\^e}nes}, {Ponthieu}, {Martin}, {Blagrave}, \&
  {Lockman}}]{Penin2011}
{P{\'e}nin}, A., {et~al.} 2011, ArXiv e-prints

\bibitem[{{Reynolds} {et~al.}(2004){Reynolds}, {Madsen}, \& {Moseley}}]{wham}
{Reynolds}, R.~J., {Madsen}, G.~J., \& {Moseley}, S.~H. 2004, \apj, 612, 1206

\bibitem[{{Schroedter} {et~al.}(2005){Schroedter}, {Badran}, {Buckley},
  {Bussons Gordo}, {Carter-Lewis}, {Duke}, {Fegan}, {Fegan}, {Finley},
  {Gillanders}, {Grube}, {Horan}, {Kenny}, {Kertzman}, {Kosack}, {Krennrich},
  {Kieda}, {Kildea}, {Lang}, {Lee}, {Moriarty}, {Quinn}, {Quinn},
  {Power-Mooney}, {Sembroski}, {Wakely}, {Vassiliev}, {Weekes}, \&
  {Zweerink}}]{Schroedter2005}
{Schroedter}, M., {et~al.} 2005, \apj, 634, 947

\bibitem[{{Tsumura} {et~al.}(2010){Tsumura}, {Battle}, {Bock}, {Cooray},
  {Hristov}, {Keating}, {Lee}, {Levenson}, {Mason}, {Matsumoto}, {Matsuura},
  {Nam}, {Renbarger}, {Sullivan}, {Suzuki}, {Wada}, \& {Zemcov}}]{Tsumura10}
{Tsumura}, K., {et~al.} 2010, \apj, 719, 394

\bibitem[{{Tsumura} {et~al.}(2011){Tsumura}, {Battle}, {Bock}, {Cooray},
  {Hristov}, {Keating}, {Lee}, {Levenson}, {Mason}, {Matsumoto}, {Matsuura},
  {Nam}, {Renbarger}, {Sullivan}, {Suzuki}, {Wada}, \& {Zemcov}}]{Tsumura11}
---. 2011, ApJS (Submitted)

\bibitem[{{Vanhollebeke} {et~al.}(2009){Vanhollebeke}, {Groenewegen}, \&
  {Girardi}}]{trilegal}
{Vanhollebeke}, E., {Groenewegen}, M.~A.~T., \& {Girardi}, L. 2009, \aap, 498,
  95

\bibitem[{{Wright}(1998)}]{Wright98}
{Wright}, E.~L. 1998, \apj, 496, 1

\bibitem[{{Wright}(2001)}]{Wright2001}
---. 2001, \apj, 553, 538

\bibitem[{{Yamamuro} {et~al.}(2006){Yamamuro}, {Sato}, {Zenno}, {Takeyama},
  {Matsuhara}, {Maeda}, \& {Matsueda}}]{Yamamuro2006}
{Yamamuro}, T., {Sato}, S., {Zenno}, T., {Takeyama}, N., {Matsuhara}, H.,
  {Maeda}, I., \& {Matsueda}, Y. 2006, Optical Engineering, 45, 083401

\bibitem[{{Zemcov} {et~al.}(2011){Zemcov}, {Arai}, {Battle}, {Bock}, {Cooray},
  {Hristov}, {Keating}, {Kim}, {Lee}, {Levenson}, {Mason}, {Matsumoto},
  {Matsuura}, {Nam}, {Renbarger}, {Sullivan}, {Suzuki}, {Tsumura}, \&
  {Wada}}]{Zemcov11}
{Zemcov}, M., {et~al.} 2011, ArXiv e-prints 1112.1424

\bibitem[{{Zemcov} {et~al.}(2010){Zemcov}, {Blain}, {Halpern}, \&
  {Levenson}}]{Zemcov2010}
{Zemcov}, M., {Blain}, A., {Halpern}, M., \& {Levenson}, L. 2010, \apj, 721,
  424

\end{thebibliography}

\end{document}